\documentclass[conference]{IEEEtran}

\usepackage[utf8]{inputenc}
\usepackage{tikz}
\usepackage{calc}
\usepackage{amsmath}
\usepackage{pgfplots}
\usepackage{amssymb}
\usepackage{listings}
\usepackage{xpatch}
\usepackage[ruled]{algorithm2e}
\usepackage{mathtools}
\usepackage{graphicx}
\usepackage{float}
\usepackage{subfig}
\usepackage[nomessages]{fp}

\usetikzlibrary{automata,positioning,arrows.meta,shapes.geometric,decorations.markings, fit}
\usetikzlibrary{calc,decorations,bayesnet}

\tikzset{>=stealth}
\usepackage[margin=1in]{geometry}
\usepackage[colorlinks,citecolor=blue,linkcolor=blue,urlcolor=blue]{hyperref}
\definecolor{light-gray}{gray}{0.7}

\usepackage[]{biblatex}
\usepackage{amsthm}

\makeatletter
\def\thmhead@plain#1#2#3{%
  \thmname{#1}\thmnumber{\@ifnotempty{#1}{ }\@upn{#2}}%
  \thmnote{ {\the\thm@headfont#3}}}
\let\thmhead\thmhead@plain
\makeatother

\newcommand{\anon}[2]{
\ifdefined\anonymous%
#1
\else 
#2%
\fi%
}

\newcommand{\platformname}{World of Bugs}
\newcommand{\acr}{WOB}
\newcommand{\mlagents}{ML-Agents}

\newcommand{\website}{https://benedictwilkins.github.io/world-of-bugs/}

\usepackage[normalem]{ulem}

\def\BibTeX{{\rm B\kern-.05em{\sc i\kern-.025em b}\kern-.08em
    T\kern-.1667em\lower.7ex\hbox{E}\kern-.125emX}}

\pgfplotsset{compat=1.17} 
\begin{document}

\title{World of Bugs: A Platform for Automated Bug Detection in 3D Video Games}

\anon{}{
\author{\IEEEauthorblockN{1\textsuperscript{st} Benedict Wilkins}
\IEEEauthorblockA{\textit{Computer Science} \\
\textit{Royal Holloway University of London}\\
London, UK \\
   0000-0002-9107-2901 }
\and
\IEEEauthorblockN{2\textsuperscript{nd} Kostas Stathis}
\IEEEauthorblockA{\textit{Computer Science} \\
\textit{Royal Holloway University of London}\\
London, UK \\ 0000-0002-9946-4037}}
}

\maketitle

\begin{abstract}
We present \platformname\ (\acr), an open platform that aims to support Automated Bug Detection (ABD) research in video games. We discuss some open problems in ABD and how they relate to the platform's design, arguing that learning-based solutions are required if further progress is to be made. The platform's key feature is a growing collection of common video game bugs that may be used for training and evaluating ABD approaches. 

\end{abstract}


\begin{IEEEkeywords} Video Games, Automated Testing, Automated Bug Detection (ABD), Machine Learning, ABD Platform
\end{IEEEkeywords}

\section{Introduction}

Video games are uniquely positioned in the space of software. They are designed to be an immersive experience for the user with a focus on narrative, interactivity and rich sensory presentation. However, the level of immersion a player experiences is drastically diminished if bugs are discovered during play. Although the industry invests significant effort in testing games before launch, it is still common for at least some bugs to slip through. One of the major contributing factors is the relative size and complexity of modern video games. It is becoming impractical to manually test every interaction and explore everything that a game has to offer even with a dedicated team of testers. 

With recent developments in automated game playing, software agents may offer a solution to the testing automation problem. Automated bug detection (ABD) is an emerging field of study that is receiving increasing attention in the video games research community \cite{Ariyurek2021, Gordillo2021, Gudmundsson2018, Albaghajati2020, Wilkins2020, Pfau2017, zheng2019}. The focus of ABD research is on designing software agents capable of (a) \textit{exploration}, playing the game in order to discover bugs, and (b) \textit{identification}, to determine whether a particular situation contains a bug.

\renewcommand\wr{.22}
\begin{figure}%
    \centering
    \vspace{-.3cm}
    \subfloat[Texture Missing\centering]{
    \includegraphics[width=\wr\textwidth]{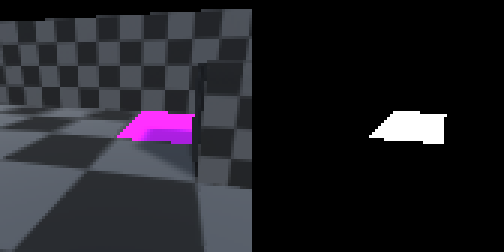}}
    \hspace{0cm} \vspace{-.3cm}
    \subfloat[Texture Corruption\centering]{
    \includegraphics[width=\wr\textwidth]{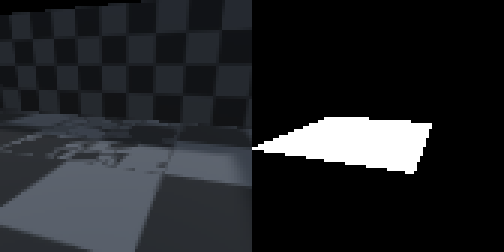}}
    \hspace{0cm} \vspace{-.3cm}
    \subfloat[Z-Fighting\centering]{
    \includegraphics[width=\wr\textwidth]{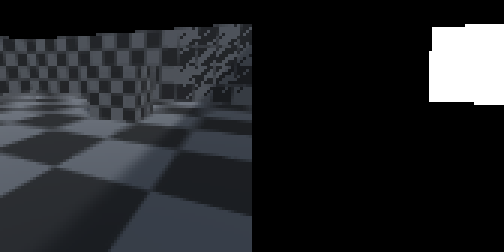}}
    \hspace{0cm} 
    \subfloat[Z-Clipping\centering]{
    \includegraphics[width=\wr\textwidth]{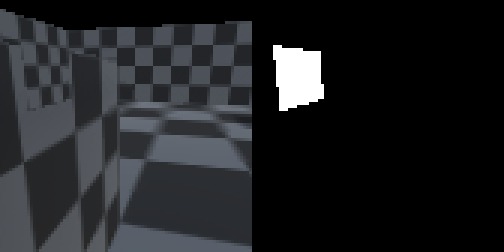}}
    \hspace{0cm} 
    \subfloat[Geometry Corruption\centering]{
    \includegraphics[width=\wr\textwidth]{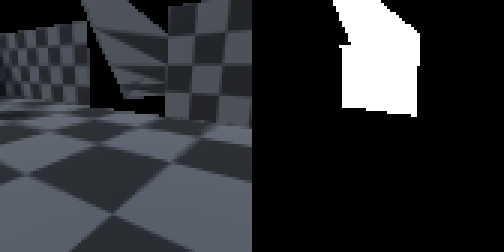}}
    \hspace{0cm} \vspace{-.3cm}
    \subfloat[Screen Tear\centering]{
    \includegraphics[width=\wr\textwidth]{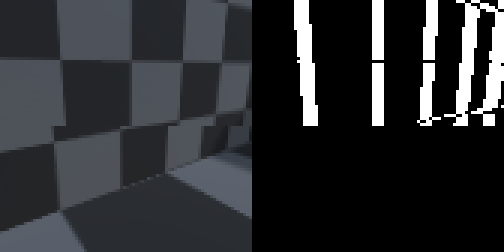}}
    \hspace{0cm} \vspace{-.3cm}
    \subfloat[Camera Clipping\centering]{
    \includegraphics[width=\wr\textwidth]{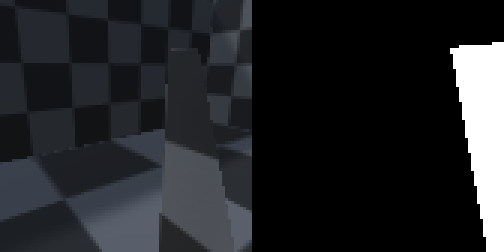}}
    \hspace{0cm} 
    \subfloat[Black Screen\centering]{
    \includegraphics[width=\wr\textwidth]{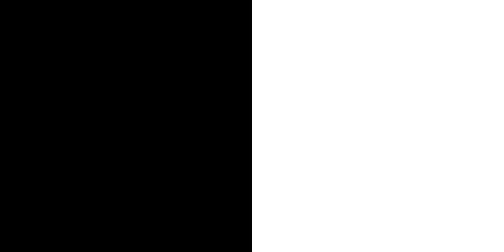}}
    \hspace{0cm} 
    \subfloat[Boundary Hole\centering]{
    \includegraphics[width=\wr\textwidth]{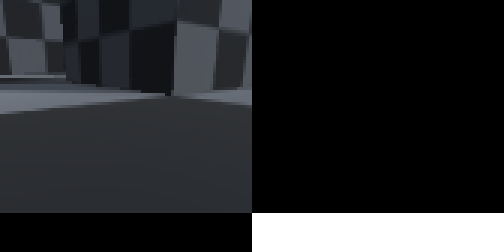}}
    \hspace{0cm}
    \subfloat[Geometry Clipping\centering]{
    \includegraphics[width=\wr\textwidth]{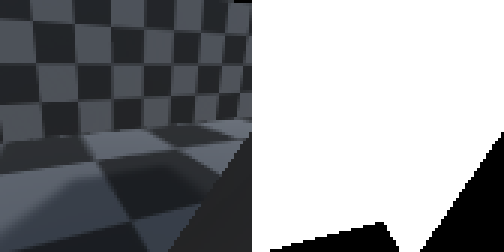}}
    \hspace{0cm}
    \caption{Some of the bugs implemented in \acr. Each agent observation (left) contains a bug which is labelled with a mask (right). }
    \label{fig:VisualBugs}
\end{figure}

As is common in an emerging field, the literature is somewhat fractured when it comes to testing and comparing approaches to these problems. Without direct ties to industry it is difficult to obtain a development version of a game, meaning that researchers need to construct their own. Ironically, as bugs are hard to come by, the experimental setup cost is high. This has discouraged progress on certain problems associated with ABD, particularly on methods for identifying some of the more challenging bugs that video games exhibit. 

With the aim of supporting further work in all aspects of ABD we have developed \platformname\ (\acr), an open experimentation platform built with the Unity3D engine. \acr\ provides a means of training and testing ABD agents in both exploration and bug identification. The core contribution of this work is in a growing collection of common video game bugs \cite{Levy2009GameDE, Lewis2010, Nantes2013} implemented in the platform, their integration with the abstractions provided by Unity's \mlagents\ package \cite{Juliani2018} and support for learning-based bug identification. The platform also provides a collection of pre-built baseline game environments, agents and datasets which researchers can use to conduct experiments. 

\begin{figure}
    \centering
    \subfloat[\centering Maze-v0]{\includegraphics[width=.24\textwidth, height=0.24\textwidth]{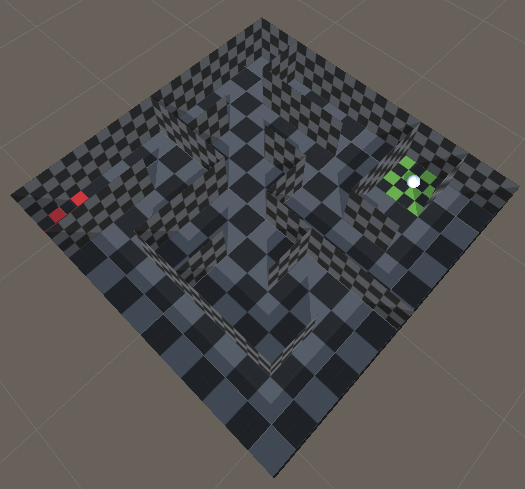}}%
    \hfill
    \subfloat[\centering
    GettingStuck-v0]{\includegraphics[width=.24\textwidth,height=0.24\textwidth]{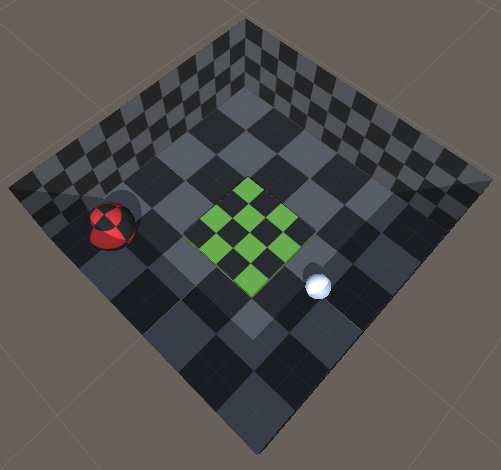}}%
    \caption{Example environments built in to \acr (from above). The agent appears as a white sphere. In (a) the agent is able to see through some walls by clipping the camera, constituting an \textit{invalid information access} bug. In (b) the agent can get stuck at the bottom of an elevator shaft after taking a particular sequence of actions. Both environments also include various perceptual bugs.}
    \label{fig:TestEnvironment}
\end{figure}

\section{\platformname\ (\acr)}
\label{sec:\platformname}



We have focused our attention on logical and perceptual bugs in 3D first person games. Logical bugs include crashes, getting stuck, performing invalid actions, out-of-bounds and invalid information access bugs, among others. Perceptual bugs manifest visually to the player (see Fig. \ref{fig:VisualBugs}) and include geometry and texture corruption, clipping issues, shadow-acne, and other rendering related issues. This setup provides a substantial challenge but is not completely out of reach of modern machine learning algorithms for both the exploration and identification problems. 

Of the approaches to ABD many focus on developing explorative agents. Works such as \cite{Bergdahl2020, Gordillo2021, zheng2019} focus on learning agents capable of finding simple logical bugs (e.g. getting stuck, or out-of-bounds) by extensive exploration of the game with rules or guards for identification. However, the sorts of bugs that game testers are looking for are often not easily identified with such methods. High-level behavioural bugs such as those related to progression, narrative, or involve complex interactions or game AI are difficult to identify with rules, and it is not yet clear how they may be tackled in general. 

For perceptual bugs and the more complex logical bugs, which have scarcely been explored in conjunction with exploration, vision and learning based methods are a promising direction \cite{Wilkins2020, Nantes2013}. Progress on these bugs is an essential next step in the development of more general ABD agents that do not rely on hand-crafted rules for identification. The primary purpose of \acr\ is to support work in this area, but it also has the potential to be extended to support research into the more challenging behavioural bugs.

Details on the implementation of bugs, the tools for developing agents and environments, and some examples are given in the next sections. The platform is available as a Unity package on the open source unity package registry OpenUPM, and on GitHub. Links and further details can be found \href{\anon{}{\website}}{here}\footnote{\anon{LINK REMOVED FOR DOUBLE BLIND REVIEW }{\href{\website}{\website}}}.




\subsection{Agents}

An agent in \acr\ is created by default with three sensors: a main camera, which renders the view of the scene as the player would see it, a bug mask camera, which renders a mask over the scene showing in which regions there is a bug present, and a sensor which records various environment/agent properties such as the agent's position and rotation. The bug mask acts as a label for the agents observation and is instrumental in enabling machine learning to be applied to the bug identification problem. In an attempt to standardise the comparison of explorative agents, \acr\ also implements a set of common actions including movement, jumping, view changes and simple event-based interaction, which can be configured during setup. Agent behaviours, goals and bug identification models may be specified in Python or in C\# with the support of \mlagents\ and associated Python packages.

\acr\ provides a simple navigational agent that uses Unity's built in navigation system inspired by \cite{Prasetya2020}. The agent's behaviour is to walk and look around, picking random points in the scene to navigate to. 
The agent can be used in simpler environments where extensive exploration is not necessary to ease experimentation with approaches to bug identification. Additionally \acr\ provides player controlled agents that may be used for debugging new environments or bug implementations, or for generating training data. The intention is that with time the community will provide a collection of ABD agents that may serve as comparisons for further work.

\subsection{Bugs}

Bugs are implemented as collections of scripts and shaders, they can be added to the scene and enabled/disabled from our Python API. Newly implemented bugs will be automatically registered with the API when added to the scene. Implementing a bug requires two considerations: how to manifest the bug in the main camera and where to mask in the mask camera. The first is straightforward, and often is just a matter of writing a script to modify some properties of the environment (e.g. deleting a texture for a missing texture bug). The second is more challenging, and for this reason we have provided some scripts and shaders to ease the process. Every bug has a tag, a unique identifier which determines its colour in the bug mask. Bug tags can be used to label bugs that are associated with particular objects in the environment, such as missing textures, corrupted geometry and bad animations. Once tagged, these objects will be rendered in the bug mask. Bugs such as screen tearing, flickering, and freezing, are slightly more difficult to label but typically involve a post processing step in the rendering pipeline to directly modify the final bug mask. For details on how this may be implemented refer to the supplementary documentation. The bug mask shader also contains code for rendering bugs that are not associated with a particular object and do not effect the whole scene, a good example is camera clipping (seeing through walls). It will automatically render the back-side geometry as part of the mask, that is, if the agent can see inside a geometry, the inside faces will be rendered in the mask. Additionally, it will render the sky box below a certain height, this may be used to label so-called \textit{map-hole} or \textit{boundary bug} (see Fig. \ref{fig:VisualBugs}). 

The choice to label bugs as part of a mask over the players observation (the final rendered screen) has been made with generality and extensibility in mind. All bugs manifest at this level in some form\footnote{Audio related issues are not considered in this work, but a similar idea could be applied there.}. In cases where the bug includes a temporal aspect, for example with a miss-behaving NPC, the mask can be extended through time. A mask provides the greatest flexibility for labelling newly implemented bugs and gives more information than a single scalar label. 

Labelling a newly implemented bug is an important design choice. The most appropriate mask is one that gives the most information about what has gone wrong. In many cases this corresponds to masking the afflicted game object, for example in the case of a misbehaving NPC the simplest choice may be to highlight its body. If a cut-scene fails to show at a given point things are less clear, perhaps the whole observation should be masked. For bugs such as the so-called \textit{invalid information access} bug \cite{Lewis2010} in which a player gains access to information they are not supposed to have, the source (or a proxy indicator) of the information can be masked. A common example that is already supported by \acr\ is \textit{camera clipping} (looking through a wall). This bug may enable the player to see the structure or exit of an upcoming maze (see Fig. \ref{fig:TestEnvironment}.a). In such a case, backside faces may be visible and act as a proxy indicator.

The label mask can be used in a number of ways, for example, to test different inductive biases for learning models, for bug segmentation or classification, or simply in model evaluation. The platform is meant only as a means to experiment with different approaches to ABD. Practical deployment of these models is currently beyond the scope of this work, but is an important consideration going forward.

\subsection{Environments}

At this time \acr\ has a number of built-in environments, details for each can be found in the platform documentation. They provide a stable testing ground for ABD approaches and have been designed either to exhibit particular bugs (see Fig. \ref{fig:TestEnvironment}), or simplify certain aspects of the ABD problem. For example, one environment consists of a single static room in which the agent is able to move and look around. This environment should generally be used to test approaches to identifying perceptual bugs as exploration is trivial. In addition to the built-in environments currently available, new game environments can be built for testing specific kinds of behaviours or to support new kinds of bugs. The tools are available to make this process relatively straight forward. The current set of implemented bugs have been designed to be as generic as possible and can be applied directly to any newly implemented 3D environments.

\subsection{Python API}

To better manage experiments and open the platform up to the Python machine learning ecosystem, \acr\ exposes a Python API which makes heavy use of \mlagents\ and OpenAI gym \cite{Brockman2016}. In addition to the interface provided by these packages, the \acr\ API allows users to enable and disable bugs, change agent behaviours and control other aspects of the environment and agents at run-time. 

\subsection{Datasets}

To our knowledge there are currently no publicly available datasets that contain examples of video game bugs that are suitable for training/testing learning-based approaches to bug identification\footnote{\cite{Wilkins2020} provides a dataset, however the bugs are artificial and unrealistic.}. As it is often easier to work with a static dataset when training learning models, \acr\ has explicit support for generating datasets given an environment and a game playing agent. We have generated a collection of datasets using the built-in agents and environments which we hope will serve as baselines for bug identification research. They consist of large (up to 500k) collections of labelled observations and actions. Links to datasets can be found in the platform documentation.

\section{Limitations}

The focus of \acr\ is currently on visual issues, we have made no consideration of audio related problems. While not explicitly supported, many of the design decisions that have gone into visual issues could be reused for audio. 
Although \mlagents\ support parallel agents running in the same instance of Unity. This is not currently supported by \acr\ as our Python API makes use of OpenAI gym which supports only single agent interactions. It is however possible to create multiple instances of any \acr\ environment running as separate processes to enable parallel training. 

\section{Conclusions \& Future Work}

The bugs implemented thus far are a small portion of the many diverse bugs that video games exhibit. There are plans to grow the list of available bugs and move beyond simpler logical and perceptual bugs into more challenging areas such as progression and narrative. For this to happen however, richer game environments are required. The game environments currently provided are simple when compared with games that ABD might eventually be applied to in practice. We hope that over time the community will provide some more interesting examples that may exhibit a wider variety of bugs. ABD has great potential in addressing some key issues in testing modern video games, but to do so requires a deeper exploration of both game playing and bug identification. \acr\ provides researchers with the tools required to make progress on some of the key problems in ABD.

\printbibliography

\end{document}